\documentclass[letterpaper,twocolumn,10pt]{article}
\IfFileExists{usenix-2020-09.sty}{%
  \usepackage{usenix-2020-09}%
}{%
  \usepackage[letterpaper,textwidth=7in,textheight=9in,centering]{geometry}
  \usepackage{times}
  \setlength{\columnsep}{0.33in}
}
\usepackage{amsmath,amssymb,mathtools}
\usepackage{booktabs}
\usepackage{multirow}
\usepackage{array}
\usepackage{tabularx}
\usepackage{enumitem}
\usepackage{xspace}
\usepackage{graphicx}
\usepackage{xcolor}
\usepackage{balance}
\usepackage{url}
\usepackage{capt-of}
\usepackage{microtype}
\IfFileExists{artifact_url.tex}{

}{}

\newcommand{\sys}{VectorHijack-SR\xspace}

\newcommand{\Dq}{D_{q}}
\newcommand{\Dtask}{D_{task}}
\newcommand{\Dshift}{D_{shift}}
\newcommand{\AGR}{\mathrm{AGR}}

\newcommand{\Cstruct}{C_{\mathrm{struct}}}
\newcommand{\Ctau}{C_{\tau}}
\newcommand{\TopK}{\mathrm{TopK}}

\newcommand{\Prb}{\mathbb{P}}

\newcolumntype{L}[1]{>{\raggedright\arraybackslash}p{#1}}
\newcolumntype{C}[1]{>{\centering\arraybackslash}p{#1}}
\newcolumntype{Y}{>{\centering\arraybackslash}X}
\newcolumntype{Z}{>{\raggedright\arraybackslash}X}
\newcommand{\TableBody}{\footnotesize\setlength{\tabcolsep}{4pt}\renewcommand{\arraystretch}{1.12}}
\newcommand{\TableBodyTight}{\footnotesize\setlength{\tabcolsep}{3pt}\renewcommand{\arraystretch}{1.12}}

\date{}
\title{\Large \bf When Do PEFT Adaptations Leak Structure?\\Measuring Black-Box Structural Bounds in Public-Base Model Services}

\author{
Zhongjiang Yao$^{2}$,
Shuangshuang Liang$^{3}$,
Chun Yang$^{1,*}$,
Liwei Chen$^{1}$,
Gang Shi$^{1}$\\
$^{1}$Institute of Information Engineering,CAS,Beijing,China\\
$^{2}$King's Collage London,UK\\
$^{3}$Weibo Corporation,Beijing,China\\
$^{*}$Corresponding author: yangchun@iie.ac.cn
}

\begin{document}
\maketitle

\begin{abstract}
Services increasingly deploy a public foundation model together with a private parameter-efficient adaptation, creating a differential leakage surface when an auditor or adversary can execute the public base locally and observe rich victim outputs. We present \sys, a measurement methodology that turns paired victim/base residuals into calibrated structural bounds over PEFT family, layer locality, and coarse rank, while explicitly separating metadata visibility from open-world validity and operational exploitability. The estimator aggregates per-query magnitude, ranking, entropy, margin, length, template, locality, and spectral statistics into service-level views; a service-disjoint closed-set classifier produces structural probabilities, and a cross-fitted hierarchical rejector tests whether the victim lies outside the calibrated LoRA manifold. Across classification backbones, family leakage exceeds uniform chance on BERT/MNLI (8/12), RoBERTa/MNLI (21/24), and DeBERTa-v3 on MNLI (12/18) and AG News (15/18), while rank evidence is task dependent: BERT/MNLI and DeBERTa/AG News reach 8/9, but DeBERTa/MNLI reaches 4/9 and is statistically compatible with chance after multiplicity correction. On a ten-seed BERT open-set grid, the hierarchical rejector reaches pooled AUROC 0.804 (95\% CI $[0.660,0.927]$) and known accuracy 0.956, yet fails on structurally close DoRA and LoRA+head variants. Exact-version linkage on five held-out LoRA-r64 services reaches AUC 0.940 and FMR 0.10 at 95\% true-match recall. The same experiments expose a visibility--exploitability gap: two-stage recovery provides no fair-budget query savings, posterior-selected PEFT is less accurate than distill-then-convert-to-PEFT (0.356 versus 0.517), and free-running generation is near chance; deterministic one-token results therefore serve only as a trajectory-alignment diagnostic. We conclude that known-base, rich-output PEFT services can leak actionable structure and private-version information, but closed-set confidence alone is not evidence of universal adapter extraction.
\end{abstract}

\section{Introduction}

Parameter-efficient fine-tuning changes what a black-box service keeps private.  In a conventional extraction problem, the unknown object is an entire function.  In a public-base PEFT service, the base checkpoint is available and the confidential component is a compact, structured delta.  This differential structure creates a new leakage surface: paired victim/base outputs may reveal how the service was adapted even when they do not reveal the private weights.

We study three questions.  \textbf{RQ1: Visibility.} Under which interfaces do residuals reveal PEFT family, locality, or a coarse rank bucket?  \textbf{RQ2: Validity.} Does a high-confidence closed-set posterior remain meaningful when the victim uses an unseen structure or when the locally executed base differs from the served base?  \textbf{RQ3: Consequence.} Does visibility enable reliable private-version linkage, query savings, or a more useful deployable artifact than strong distillation baselines?

\sys addresses RQ1 by transforming paired victim/base responses into residual magnitude, ranking, entropy, locality, template-sensitivity, and spectral signatures.  A calibrated posterior returns a structural \emph{bound}, not a claim of unique weight or rank recovery.  We address RQ2 with disjoint open-set splits and real base perturbations, and RQ3 with held-out checkpoint fingerprinting and cost-matched recovery.  This claim ladder separates metadata leakage, lineage leakage, and artifact recovery; evidence at one level is not promoted to the next.

The results identify a conditional but security-relevant phenomenon.  Family-level structure is visible on several classification backbones, while rank evidence varies sharply by task.  We quantify these small-sample results with Wilson intervals and prespecified exact binomial tests, correcting the primary family/rank comparisons with Holm's procedure.  This analysis matters: DeBERTa/MNLI family inference is significant, but its 4/9 rank result is statistically compatible with chance.  Open-set rejection improves substantially on an expanded ten-seed BERT grid, yet structurally close DoRA and LoRA+head variants remain indistinguishable.  Modest quantization and continued-pretraining drift preserve BERT family accuracy through a residual ratio of 1.19, but the mismatch detector itself remains near chance, making exact base alignment an auditable prerequisite rather than an assumption an external attacker can verify.

The strongest operational consequence is private-version linkage.  With task, base, family, target modules, and rank held fixed, residual signatures distinguish the same checkpoint from independently trained checkpoints with AUC 0.940 and FMR 0.10 at 95\% recall on five held-out services.  The remaining consequence tests are deliberately less positive.  Structural probing does not save queries under fair accounting, free-running generation hides the signal, and posterior-selected PEFT does not beat distill$\rightarrow$PEFT on fidelity.  The recovery experiment instead exposes a cost--format frontier: the posterior route produces a 10.2\,MB attachable delta at 1.00$\times$ training cost, whereas distill$\rightarrow$PEFT reaches higher accuracy at 4.46$\times$ cost.

These results support a measurement contribution rather than a universal commercial-API attack.  Label-only interfaces are near chance for the present passive method.  Deterministic short-answer decoder endpoints are useful for explaining trajectory alignment, but they are not treated as representative conversational APIs.  The practical target is a known-base, rich-output setting such as internal governance, platform onboarding, model provenance checks, or an authorized red-team interface.

We make four contributions.
\begin{itemize}[leftmargin=*]
    \item We formalize PEFT structural leakage and introduce \sys, a residual-signature methodology that reports posterior bounds over family, locality, and rank buckets.
    \item We provide statistically qualified multi-backbone measurements, including Wilson intervals, exact chance tests, multiplicity correction, teacher-forcing controls, and fair query accounting.
    \item We evaluate two prerequisite gates that closed-set studies typically omit: open-world structure rejection and public-base mismatch sensitivity.  The results identify both detectable off-manifold shifts and unresolved near-neighbor failures.
    \item We measure the visibility--exploitability gap through held-out private-version linkage and cost-matched recovery.  Linkage is the strongest positive consequence; recovery yields a deployment-cost trade-off rather than fidelity or query superiority.
\end{itemize}
\section{Problem Setting}
\label{sec:problem}

\subsection{Deployment model}

A victim service consists of a public base model $f_b(\cdot;W_b)$ and a private adaptation $\theta^\star$.  The deployed model is
\begin{equation}
    f_v(x) = g(x; W_b, \theta^\star).
\end{equation}
The private adaptation belongs to a restricted family: LoRA-style low-rank updates, adapter blocks, prefix or prompt parameters, task heads, or localized top-layer updates.  For LoRA, a target linear layer is modified as
\begin{equation}
    W' = W + A B^\top, \qquad A \in \mathbb{R}^{d\times r},\; B \in \mathbb{R}^{k\times r}.
\end{equation}
We do not assume that the exact private matrices are identifiable.  LoRA matrices have reparameterization freedom, and many PEFT objects can be behaviorally equivalent on a finite distribution.

\subsection{Threat model and attacker objectives}
\label{sec:threat}

\paragraph{Primary setting: known-base structural audit.}
We primarily study a \emph{known-base auditor}---a service owner, platform compliance checker, or authorized red team---who knows the exact public base checkpoint, tokenizer, and inference code, and can execute that base locally.  The auditor can submit adaptively chosen inputs to a PEFT-backed service and observe only the interface outputs defined below.  The auditor does not know the private PEFT weights, private fine-tuning data, optimizer state, random seed, server-side activations, or deployment metadata.  Query budgets include every victim call used for calibration, structural inference, fingerprinting, and recovery.  This matched-base assumption defines the primary measurement condition rather than an implicit guarantee.

This framing matches internal PEFT governance and research APIs that expose logits or top-$k$ scores ($C1$--$C3$, $G1$--$G3$).  Section~\ref{sec:findings} shows that passive residual signatures collapse to chance on label-only ($C4$) interfaces; active decision-boundary probing is outside the present method.

\paragraph{Matched base as a risk, not a free assumption.}
We separately evaluate quantization, checkpoint revision, tokenizer or chat-template drift, hidden system prompts, decoding-policy mismatch, and continued-pretraining drift.  When the auditor's local base diverges from the served base, structural conclusions become attacker-side uncertainty: a near-chance mismatch detector does not grant the auditor a free pass, but it also means an external adversary cannot trust structural bounds without verifying base alignment.  We therefore report a \emph{robustness boundary}---family/rank accuracy as a function of residual-ratio severity---rather than claiming universal mismatch detection.

\paragraph{Assets and goals.}
The defender's confidential assets include: (i) the PEFT family and target modules, (ii) rank or bottleneck allocation, (iii) the identity or lineage of a private adapter checkpoint, and (iv) a compact adaptation artifact that can be redeployed with the public base.  The attack (or audit) has three increasingly strong goals.  \emph{G1: structural bounding} narrows the candidate set over family, locality, and rank bucket.  \emph{G2: adapter fingerprinting} links two services or attributes a service to a known private adaptation lineage.  \emph{G3: redeployable recovery} produces a PEFT-format substitute that can be loaded on the public base and evaluated for direct weight merge and adapter composition.

We explicitly exclude white-box memory disclosure, compromise of the serving host, access to gradients or hidden states, and unique recovery of LoRA factors $A$ and $B$.  Because $AB^\top=(AR)(BR^{-\top})^\top$ for any invertible $R$, factor-level equality is neither identifiable nor required.  The correct comparison is therefore between operational artifacts: a behavior-only student and a compact base-compatible adaptation.

\subsection{Structural leakage}

Let $O_v(x)$ be the victim output under interface $I$ and $O_b(x)$ the public-base output.  A structural hypothesis is
\begin{equation}
    h=(\tau,S,\rho)\in\mathcal{H},
\end{equation}
where $\tau$ is a PEFT family, $S$ is a layer or module set, and $\rho$ is a rank, bottleneck, or related hyperparameter.  The structural-inference target is the posterior
\begin{equation}
    p(h\mid Q_{struct}).
\end{equation}
The output of inference is a bound: top-$k$ hypotheses, bucket labels, posterior entropy, and the number of queries required to reach a confidence threshold.

\subsection{Three-gate observability model}
\label{sec:three_gate}

Closed-set accuracy is only one component of end-to-end leakage. Let $A$ denote sufficient alignment between the local and served base, $M$ denote that the private adaptation is either represented by or detectably outside the calibrated structure manifold, $I$ denote an interface that preserves the residual signal, and $C$ denote success of the requested consequence once the first three conditions hold. Defining end-to-end success as $S=A\cap M\cap I\cap C$, the chain rule gives
\begin{equation}
\Prb(S)=\Prb(A)\Prb(M\mid A)\Prb(I\mid A,M)\Prb(C\mid A,M,I).
\end{equation}
The base-mismatch study evaluates the alignment gate, open-set rejection the manifold gate, interface and trajectory experiments the observation gate, and fingerprinting/recovery the conditional consequence. This decomposition is an accounting device, not an independence assumption, and prevents a strong closed-set classifier from being interpreted as an end-to-end attack probability.

\subsection{Behavioral recovery}

For a fixed structure $h$, behavioral recovery solves
\begin{equation}
    \hat{\theta}_{h} = \arg\min_{\theta\in\Theta(h)} \sum_{x\in Q_{param}} L\big(O_v(x), O_{\theta,h}(x)\big) + \lambda \Omega(\theta).
\end{equation}
We use this objective only to test whether inferred structure has operational value.  Recovery is evaluated operationally; factor-level identity is outside the estimand.

\subsection{Behavioral parity versus operational equivalence}

A plain distillation student and a PEFT substitute may attain similar agreement while remaining operationally different.  We represent an extracted artifact by the capability vector
\begin{equation}
\begin{aligned}
\mathcal{U}(m)=(&\AGR,\mathrm{Params},\mathrm{Attach},\\
                &\mathrm{Merge},\mathrm{Compose},\mathrm{Transfer}).
\end{aligned}
\end{equation}
Here, \emph{Attach} tests whether the artifact loads on the public base without architectural surgery; \emph{Merge} tests whether its delta can be folded into the base checkpoint; and \emph{Compose} measures retained utility after combining it with a second independently trained adapter.  Distillation is a strong baseline for $\AGR$, but it is structurally incapable of satisfying the PEFT-format dimensions without an additional conversion or retraining step.

\subsection{Interfaces and data splits}

For classification, $C1$ returns full logits or probabilities, $C2$ returns top-$k$ logits, $C3$ returns a selected score, and $C4$ returns labels only.  For generation, $G1$ returns token-level logits, $G2$ returns top-$k$ token scores, $G3$ returns sequence scores, and $G4$ returns final text only.

We use three distributions.  $\Dq$ is the adversary's query distribution.  $\Dtask$ is the held-out task distribution.  $\Dshift$ contains template, lexical, length, or neighboring-task shifts and is used only for evaluation.

\subsection{System Overview}
\label{sec:overview}

\begin{figure*}[t]
\centering
\includegraphics[width=0.95\textwidth]{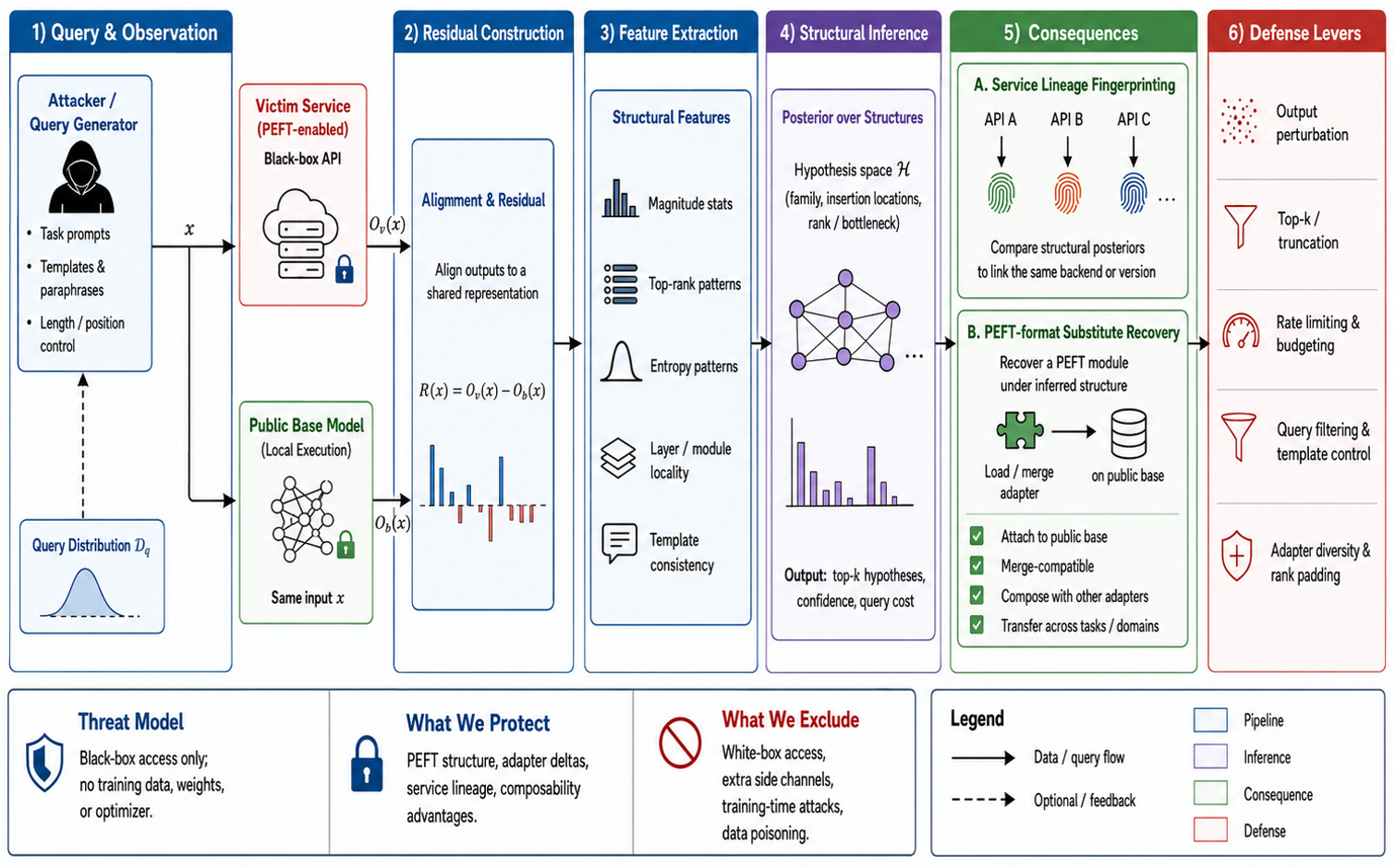}
\caption{System overview of \sys. Identical probes are sent to the PEFT-enabled service and the locally executed public base. Aligned residual signatures feed a calibrated structural estimator, an independently thresholded unknown/base-alignment gate, and two consequence paths: held-out private-version linkage and PEFT-format recovery. Output and query controls act before residual observation.}
\label{fig:system_overview}
\end{figure*}

Figure~\ref{fig:system_overview} summarizes the attacker--defender interaction.  The attacker sends the same probe to the remote victim and the locally executed public base, aligns the observable outputs, and converts their residual into magnitude, ranking, entropy, locality, template, and spectral features.  A calibrated posterior produces a structural bound.  That bound feeds two consequence paths: service-lineage fingerprinting and PEFT-format substitute training.  The defender acts only at the API observation and query-policy layers, allowing the same diagram to expose the utility--leakage trade-off.

\section{Measurement Methodology}
\label{sec:method}

This section specifies the complete decision pipeline used by the reported results. The main text contains the estimand, feature construction, fitted models, split discipline, rejection rule, and success criteria; Appendix~\ref{app:method_details} contains only numerical hyperparameters and implementation details, while Appendix~\ref{app:protocols} gives condition-specific seed identities and ablations.

\subsection{Analysis unit and pipeline}
\label{sec:analysis_unit}

For every probe $x$, the service and locally executed base receive identical serialized input. The implementation first computes per-query residual features, then forms fixed-size signature batches, and finally averages all batches belonging to one $(\text{service checkpoint},\text{probe seed})$ pair into a \emph{service view}. A service view, rather than an individual query or internal mini-batch, is the unit passed to structural, open-set, and lineage models. All views from one private checkpoint remain in the same calibration, threshold, or test partition. This prevents probe-level pseudo-replication and protocol metadata leakage.

The end-to-end pipeline has five steps: (1) align victim and base outputs under the selected interface; (2) construct a service-level residual signature; (3) estimate a closed-set structural distribution; (4) apply the independently calibrated base-alignment and unknown-structure gates; and (5) evaluate lineage linkage or PEFT-format recovery without reusing test services for calibration.

\subsection{Residual and spectral signatures}
\label{sec:signature}

For each probe, the observable residual is
\begin{equation}
    r(x)=O_v(x)-O_b(x),
\end{equation}
where the subtraction is defined only over information exposed by the interface. Full-logit interfaces use aligned vectors; top-$k$ interfaces use the union of exposed indices plus missing-value masks; label/text-only interfaces do not reconstruct hidden scores.

The per-query vector $f(x)$ contains residual $L_1/L_2/L_\infty$ magnitude, mean, variance, skewness, and kurtosis; victim/base entropy and entropy difference; top-1 agreement and top-$k$ overlap; margin difference; logit cosine similarity; input-length and tokenization-gap features; and, where available, local-perturbation and template-sensitivity terms. For a signature batch $Q$, each scalar feature is summarized by
\begin{equation}
T_j(Q)=[\mathrm{mean},\mathrm{std},q_{10},q_{25},q_{50},q_{75},q_{90}]\{f_j(x):x\in Q\}.
\end{equation}
Let $R_Q$ be the matrix whose rows are centered residual vectors. We append its first five singular values, effective rank, spectral entropy, and the number of components containing 90\% of residual energy. The resulting batch signature $z(Q)$ is averaged across batches to obtain the service view $z_s$. Only residual-derived feature families are whitelisted; task IDs, seeds, mismatch-condition IDs, and other experiment metadata cannot enter the classifier.

\subsection{Closed-set structural estimator}
\label{sec:closed_estimator}

Calibration-only means and variances standardize each service view to $\tilde z$. The implemented closed-set estimator is a class-balanced multinomial logistic model,
\begin{equation}
 p_{cl}(h\mid z)=\frac{\exp(w_h^\top \tilde z+b_h)}{\sum_{h'\in\mathcal H_{seen}}\exp(w_{h'}^\top \tilde z+b_{h'})}.
\end{equation}
We use ``posterior'' as shorthand for this calibrated discriminative probability; it is not a Bayesian posterior over private weights. For support diagnostics, each known structure also has calibration centroid $\mu_h$ and Ledoit--Wolf shrinkage precision $\Lambda_h$, giving
\begin{equation}
 d_h(z)=\sqrt{(\tilde z-\mu_h)^\top\Lambda_h(\tilde z-\mu_h)}.
\end{equation}
The reported structural bound consists of $\TopK(p_{cl})$, family/locality/rank buckets, entropy $H(p_{cl})$, and
\begin{equation}
 \Cstruct(\alpha)=\min\{|Q|:\max_h p_{cl}(h\mid z(Q))\geq\alpha\}.
\end{equation}
Rank is intentionally coarse: low ($r\leq16$), mid ($16<r\leq128$), high ($r>128$), or non-low-rank.

\subsection{Open-set rejection and base-alignment gate}
\label{sec:open_method}

A high closed-set score is not accepted without a separate support test. From $p_{cl}$ and $d_h$, we construct
\begin{equation}
 u(z)=[1-p_{max},H(p_{cl}),-m,d_{min},d_{pred},1-p_{conf},E,d_1/d_2],
\end{equation}
where $m$ is the top-two probability margin, $p_{conf}$ is a class-conditional conformal support value, $E$ is discriminative energy, and $d_1/d_2$ is the nearest-to-second-nearest distance ratio. A logistic meta-rejector is trained on \emph{leave-one-known-structure-out} pseudo-unknown services. Its known examples are generated by service-group cross-fitting; consequently, neither the closed classifier nor the rejector scores a checkpoint it used for fitting.

The primary ten-seed experiment uses a hierarchical rule. Let $s_{meta}(z)$ be the meta-rejector score and $s_{mah}(z)=\min_h d_h(z)$. Thresholds $\gamma_{meta}$ and $\gamma_{mah}$ are locked on threshold services, each receiving half of the prespecified 5\% known-reject budget:
\begin{equation}
 \mathrm{Reject}(z)=\mathbf 1[s_{meta}(z)>\gamma_{meta}\;\lor\;s_{mah}(z)>\gamma_{mah}].
\end{equation}
Real unknown families, ranks, mixtures, and target-module patterns are absent from both fitting and threshold selection. We report raw false concentration before rejection and accepted-unknown false concentration after rejection.

Base alignment is evaluated independently by fixing the private adapter and changing only the auditor's local base. For condition $c$, the severity statistic is
\begin{equation}
 \rho_c=\frac{\operatorname{median}_{x}\|O_v(x)-O_{b,c}(x)\|_2}{\operatorname{median}_{x}\|O_v(x)-O_{b,match}(x)\|_2}.
\end{equation}
A label-free mismatch classifier is calibrated on service-level signatures, but structural family/rank accuracy is always reported as a function of $\rho_c$ even when mismatch detection fails. Thus, the method exposes rather than hides the fact that an external attacker cannot reliably verify the known-base premise.

\subsection{Private-version linkage}
\label{sec:fingerprint_method}

For service views $a$ and $b$, the pair representation is
\begin{equation}
 q(a,b)=[|a-b|,(a-b)^{\odot2},a\odot b,\cos(a,b),\|a-b\|_2].
\end{equation}
A class-balanced logistic scorer is trained only on exact-checkpoint positive pairs and the strict hard negative: same task, base, PEFT family, target modules, and rank, but independently trained checkpoints. Positive pairs use distinct probe views of the same checkpoint. Calibration, operating-point selection, and testing use disjoint victim-training checkpoints. The FMR threshold is selected on validation services at 95\% true-match recall and then frozen for the test services. AUC confidence intervals use a two-way service-checkpoint bootstrap so that the quadratic number of pairs is not treated as independent evidence.

\subsection{Recovery and operational comparison}
\label{sec:recovery_method}

Stage~II trains a substitute under the Stage~I structural choice. We compare posterior-selected, oracle-structure, and wrong-structure PEFT; plain, parameter-matched, and compute-matched students; and distillation followed by PEFT conversion. All victim calls used by Stage~I and Stage~II count toward the same budget. For threshold $\tau$,
\begin{equation}
 \Ctau(m)=\min\{B:\AGR_m(B)\geq\tau\}.
\end{equation}
We separate fidelity from format economics. For artifact $m$, the operational vector is
\begin{equation}
\begin{aligned}
\mathcal F(m)=(&\mathrm{Acc},\AGR,\mathrm{Size},\mathrm{ServeMem},\\
               &\mathrm{TrainCost},\mathrm{Attach},\mathrm{Merge},\mathrm{Compose}).
\end{aligned}
\end{equation}
A fidelity advantage is claimed only when the paired confidence interval over independent runs excludes zero; attachability or a shape-correct merge is never counted as a fidelity win. A cost--format advantage is reported separately when two methods produce the same deployable format but differ in measured training cost, storage, or serving memory. This distinction is essential for interpreting the posterior-PEFT versus distill$\rightarrow$PEFT result.

\subsection{Defensive measurements}

Defenses are evaluated against the same gates rather than only against substitute accuracy. We measure benign utility, structural entropy and $\Cstruct@0.8$, unknown recall and false concentration, lineage FMR, and recovery fidelity under output rounding, top-$k$ suppression, label/text-only fallback, and query controls. Output-equivalent transformations are treated as ineffective if residual signatures and leakage metrics remain unchanged.

\section{Experimental Setup}
\label{sec:setup}

\subsection{Evaluation matrix}

Table~\ref{tab:matrix} separates primary evidence from corroboration and mechanism diagnostics.  Proxy transfer checks are retained only in the artifact and are not pooled into primary confidence intervals.

\begin{table*}[t]
\centering
\caption{Evaluation matrix and evidence tier. ``Primary'' rows support paper claims; corroboration tests backbone/task transfer; diagnostics explain mechanisms but are not treated as deployment-wide attack evidence.}
\label{tab:matrix}
\TableBody
\begin{tabularx}{\textwidth}{@{}L{1.05in}L{1.15in}Z L{1.10in}L{1.05in}L{1.15in}@{}}
\toprule
Evidence tier & Base/task & Victims & Interface & Statistical unit & Role \\
\midrule
Primary closed set & BERT/MNLI & Base, LoRA-r8/r64/r256, FFT & Full logits; top-$k$; label & 12 victim/seed outcomes & visibility and fair-budget recovery \\
Corroboration & RoBERTa/MNLI, QQP, SST-2 & LoRA, adapters, head, prefix & Full logits; top-$k$; label & 24 MNLI + 6 cross-task outcomes & family/task transfer \\
Corroboration & DeBERTa-v3/MNLI, AG News & LoRA, adapters, head, BitFit & Full logits; top-$k$; label & 18 outcomes per task & modern discriminative backbone \\
Mechanism diagnostic & Llama-3.1-8B-Instruct & LoRA-r8/r64/r256, top-layer FT & Constrained token logits & 36 family / 27 rank outcomes & trajectory alignment only \\
Primary open world & BERT/MNLI & 10 training seeds; unseen ranks/families/targets & Full logits & service-cluster bootstrap & unknown rejection \\
Primary consequence & BERT/MNLI LoRA-r64 & disjoint train/validation/test checkpoints & Full logits & five held-out services & exact-version linkage \\
Primary recovery & BERT/MNLI & eight cost/query-matched artifacts & Full logits & paired seed-level differences & cost--format frontier \\
Defensive implication & BERT/MNLI, LLaMA/Alpaca & output projections and query controls & Full logits/Token logits plus weak views & descriptive & no standalone defense claim \\
\bottomrule
\end{tabularx}
\end{table*}

\subsection{Modern-model corroboration}
\label{sec:modern_models}

DeBERTa-v3 prevents the classification result from being interpreted solely as a BERT-era artifact~\cite{he2021debertav3}.  The effect is task dependent: family inference is above chance on both tasks, but rank inference is statistically supported on AG News and not on MNLI (Table~\ref{tab:rank_stats}).  Llama-3.1 one-token classification is reported separately as a shared-prefix mechanism diagnostic~\cite{meta2024llama3}; it is not pooled with the primary classification evidence.

\subsection{Induced trajectory alignment for generation}
\label{sec:induced_alignment}

The free-running failure does not establish that generative PEFT structure is intrinsically hidden.  We therefore add an attack variant that deliberately suppresses trajectory divergence.  Each semantic task is rewritten as: (i) a single-token multiple-choice answer from a fixed label set, (ii) a yes/no answer with $\texttt{max\_new\_tokens}=1$, or (iii) a two-token answer-only template.  Decoding uses temperature zero and no sampling.  The attacker compares victim and base next-token distributions at the shared prompt boundary, so both models are conditioned on an identical history.

The decisive comparison is free-running text versus constrained one-token output at equal query budget.  A recovery of family or rank-bucket accuracy under constrained prompts would turn the current negative result into a concrete defense insight: output-format freedom and long stochastic trajectories suppress structural observability, whereas deterministic short-answer endpoints re-open it.  Failure under both modes would support the stronger claim that the tested decoder-side structures are genuinely hard to distinguish through outputs.

\subsection{Probe construction}

We use ID probes, OOD probes, template probes, and local perturbation probes.  The completed quick checks use approximately 1.4--1.5k probes after filtering perturbations without synonym hits.  The extended matrix keeps the same probe taxonomy so that posterior concentration can be compared across tasks and interfaces.

\subsection{Metrics and statistical analysis}

Structural metrics are family top-1/Hit@$k$, rank-bucket accuracy, posterior entropy, and $\Cstruct@0.8/0.9$.  Recovery uses agreement rate (AGR), behavior fidelity, KL divergence, and total victim calls including Stage~I.  Open-set evaluation adds AUROC, AUPR, OSCR, TPR at 5\% FPR, and false posterior concentration.  Lineage evaluation uses service-pair AUC, macro attribution, EER, and FMR at 95\% true-match recall.  Operational recovery reports task utility, artifact size, serving memory, measured training cost, safe-merge fidelity, and paired differences.

For the small closed-set grids, the analysis unit is a reported victim/seed outcome.  We give two-sided Wilson 95\% intervals for hit rates and prespecified one-sided exact binomial tests against uniform chance ($1/4$ for family and $1/3$ for LoRA rank bucket).  Holm correction is applied jointly to the eight primary family/rank endpoints in Tables~\ref{tab:family_stats} and~\ref{tab:rank_stats}.  Because several outcomes share a trained victim and vary only the probe seed, these intervals quantify observed-run uncertainty rather than population-wide training uncertainty.  Open-set and lineage confidence intervals instead use service/checkpoint-cluster bootstrap resampling; no probe-level pseudo-replication is used.

\subsection{Open-world and deployment-robustness experiments}
\label{sec:validity_extensions}

The primary BERT/MNLI extension uses disjoint detector-fit, threshold, and test training seeds.  It evaluates unseen ranks, DoRA, IA$^3$, LoRA+head, nonuniform layer ranks, and alternate target modules.  A separate fixed-adapter experiment perturbs only the locally executed base.  Held-out fingerprinting fixes task, base, family, target modules, and rank while changing the private checkpoint.  Cost-matched recovery charges all query, conversion, and training stages.  Non-BERT proxy checks are not pooled with these primary results.

\subsection{Implementation audits}

Three audits precede interpretation.  Base-only residuals must be zero up to numerical precision; LoRA residuals must be nonzero.  Teacher-forcing and free-running generation are separated to avoid comparing logits conditioned on different histories.  Interface experiments are implemented at the observation layer: $C2/G2$ use top-$k$ union residuals and missing masks, and $C4/G4$ use only label or text-level features.

\section{Findings}
\label{sec:findings}

\subsection{Finding 1: family leakage is repeatable, but uncertainty matters}

Table~\ref{tab:family_stats} reports family inference with uncertainty.  BERT, RoBERTa, and DeBERTa family results remain significant after Holm correction.  The wide intervals nevertheless show that the small grids estimate ranges, not precise deployment-wide rates.  Free-running generation is only descriptive and is not statistically separated from the four-way baseline.

\begin{table*}[t]
\centering
\caption{Family top-1 results with Wilson 95\% confidence intervals. $q$ is the Holm-adjusted one-sided exact binomial $p$-value against uniform four-way chance (0.25) for the eight prespecified classification family/rank endpoints. Cross-task and decoder diagnostics are reported without multiplicity-adjusted claims.}
\label{tab:family_stats}
\TableBody
\begin{tabularx}{\textwidth}{@{}L{1.60in}C{0.45in}C{0.70in}C{1.10in}C{0.70in}Z@{}}
\toprule
Setting & $n$ & Correct & Wilson 95\% CI & $q$ & Interpretation \\
\midrule
BERT/MNLI C1 & 12 & 8/12 & $[0.391,0.862]$ & 0.0083 & primary family leakage \\
RoBERTa/MNLI C1 & 24 & 21/24 & $[0.690,0.957]$ & $<2\times10^{-9}$ & strongest multi-family corroboration \\
RoBERTa QQP/SST-2 C1 & 6 & 6/6 & $[0.610,1.000]$ & -- & cross-task diagnostic; small $n$ \\
DeBERTa-v3/MNLI C1 & 18 & 12/18 & $[0.437,0.837]$ & 0.0014 & family signal, task dependent \\
DeBERTa-v3/AG News C1 & 18 & 15/18 & $[0.608,0.942]$ & $2.4\times10^{-6}$ & family signal transfers \\
LLaMA/Alpaca free running & 12 & 6/12 & $[0.254,0.746]$ & -- & mechanism boundary; not primary evidence \\
Llama-3.1 one-token & 36 & 36/36 & $[0.904,1.000]$ & -- & shared-prefix mechanism diagnostic \\
\bottomrule
\end{tabularx}
\end{table*}

\subsection{Finding 2: rank leakage is task dependent and should remain bucketed}

Table~\ref{tab:rank_stats} changes the interpretation of several raw counts.  BERT/MNLI and DeBERTa/AG News rank results exceed the three-way baseline after correction.  RoBERTa/MNLI and DeBERTa/MNLI do not; in particular, 4/9 on DeBERTa/MNLI is compatible with chance ($q=0.350$).  We therefore report low/mid/high buckets and avoid a cross-backbone claim of reliable rank recovery.

\begin{table*}[t]
\centering
\caption{LoRA rank-bucket results with Wilson 95\% intervals. The null is uniform three-way chance (1/3); $q$ uses the same Holm family as Table~\ref{tab:family_stats}.}
\label{tab:rank_stats}
\TableBody
\begin{tabularx}{\textwidth}{@{}L{1.70in}C{0.45in}C{0.70in}C{1.10in}C{0.70in}Z@{}}
\toprule
Setting & $n$ & Correct & Wilson 95\% CI & $q$ & Interpretation \\
\midrule
BERT/MNLI C1 & 9 & 8/9 & $[0.565,0.980]$ & 0.0048 & supported coarse rank leakage \\
RoBERTa/MNLI C1 & 12 & 7/12 & $[0.320,0.807]$ & 0.133 & not significant after correction \\
RoBERTa QQP/SST-2 C1 & 6 & 6/6 & $[0.610,1.000]$ & -- & cross-task diagnostic; small $n$ \\
DeBERTa-v3/MNLI C1 & 9 & 4/9 & $[0.189,0.733]$ & 0.350 & statistically compatible with chance \\
DeBERTa-v3/AG News C1 & 9 & 8/9 & $[0.565,0.980]$ & 0.0048 & supported on this task \\
Llama-3.1 one-token & 27 & 27/27 & $[0.875,1.000]$ & -- & mechanism diagnostic, not API-wide evidence \\
\bottomrule
\end{tabularx}
\end{table*}

\subsection{Finding 3: generation contains residuals, but free-running decoding hides structure}

Table~\ref{tab:a0} shows that generation failures are not caused by absent residuals.  Base-only residuals vanish, while LoRA residuals are large across seeds.

\begin{table}[t]
\centering
\caption{Generation tensor audit. Base-only residuals vanish; LoRA residuals are substantial.}
\label{tab:a0}
\TableBody
\begin{tabular}{@{}lrrrr@{}}
\toprule
Seed & Rows & Base $\Delta L_2$ & LoRA $\Delta L_2$ & Pass \\
\midrule
7   & 12000 & 0.0000 & 622.7 & yes \\
42  & 12000 & 0.0000 & 641.5 & yes \\
123 & 12000 & 0.0000 & 610.4 & yes \\
\bottomrule
\end{tabular}
\end{table}

Table~\ref{tab:tf} compares teacher forcing and free running at $T=8$.  Free-running KL is roughly 20--30 times larger than teacher-forced KL.  Structural signal is therefore entangled with trajectory drift.  Longer observation does not fix this under the current features: Table~\ref{tab:genT} shows that family and rank-bucket accuracy do not improve with $T\geq8$.

\begin{table}[t]
\centering
\caption{Teacher-forcing versus free-running generation for LoRA victims at $T=8$. Values are averaged over the three LoRA ranks per seed.}
\label{tab:tf}
\TableBody
\begin{tabular}{@{}lrrrr@{}}
\toprule
Seed & FR $\Delta L_2$ & TF $\Delta L_2$ & FR KL & TF KL \\
\midrule
7   & 868.9 & 371.0 & 12.50 & 0.622 \\
42  & 751.3 & 392.9 & 13.17 & 0.385 \\
123 & 805.7 & 365.2 & 12.08 & 0.574 \\
\bottomrule
\end{tabular}
\end{table}

\begin{table}[t]
\centering
\caption{Generation family and rank-bucket accuracy versus maximum generation length. Four-way family random accuracy is 0.25; three-way LoRA rank-bucket random accuracy is 0.333.}
\label{tab:genT}
\TableBody
\begin{tabularx}{\columnwidth}{@{}l c Y Y@{}}
\toprule
Seed & $T$ & Family acc. & Rank-bucket acc. \\
\midrule
7   & 2  & 0.750 & 0.667 \\
7   & 8  & 0.500 & 0.333 \\
7   & 16 & 0.500 & 0.333 \\
7   & 32 & 0.250 & 0.333 \\
42  & 2--32 & 0.250 & 0.333 \\
123 & 2--32 & 0.250 & 0.333 \\
\bottomrule
\end{tabularx}
\end{table}

\begin{figure}[t]
\centering
\includegraphics[width=\columnwidth]{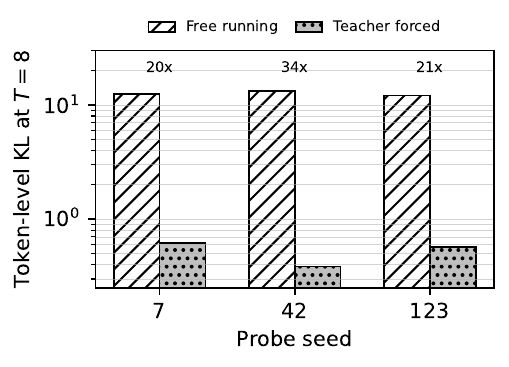}
\caption{Generation mechanism analysis: step-wise KL under teacher forcing versus free running; top-$k$ overlap at $T{=}8$ (divergence onset); early ($T{=}2$) versus late ($T\geq16$) family accuracy.}
\label{fig:gen_mechanism}
\end{figure}

\subsubsection{Constrained-generation result}

Table~\ref{tab:aligned_generation} tests whether prompt-induced alignment recovers the structural signal.  Both constrained modes outperform the matched free-running baseline across probe seeds.

\begin{table}[t]
\centering
\caption{Induced trajectory alignment on Llama-3.1-8B-Instruct. All entries are three-probe-seed aggregates at matched probe count.}
\label{tab:aligned_generation}
\TableBody
\begin{tabularx}{\columnwidth}{@{}Z C{0.48in}C{0.58in}C{0.58in}C{0.62in}@{}}
\toprule
Observation mode & Logit pos. & Family acc. & Rank acc. & $\Cstruct@0.8$ \\
\midrule
Free-running & 16/32 & 0.500 & 0.333 & 150 \\
Answer-only & 2 & 1.000 & 1.000 & 100 \\
Single label token & 1 & 1.000 & 1.000 & 100 \\
\bottomrule
\end{tabularx}
\end{table}

\subsection{Finding 4: structural bounds help avoid wrong structures, but do not beat distillation under fair budgets}

Table~\ref{tab:stage2max} shows max-budget classification recovery.  Two-stage recovery is better than wrong-structure recovery and slightly above plain distillation on mean AGR.  The margins over single-stage and distillation are small.  The result supports the claim that wrong structures hurt; it does not support a strong recovery advantage.

\begin{table}[t]
\centering
\caption{Max-budget classification recovery on MNLI, mean$\pm$std over three seeds.}
\label{tab:stage2max}
\TableBody
\begin{tabularx}{\columnwidth}{@{}Z c c c@{}}
\toprule
Attack & AGR & BF & KL$_{avg}$ \\
\midrule
Two-stage & 0.811$\pm$.015 & 0.811$\pm$.015 & 0.034$\pm$.005 \\
Single-stage & 0.805$\pm$.007 & 0.805$\pm$.007 & 0.037$\pm$.006 \\
Wrong-structure & 0.774$\pm$.002 & 0.774$\pm$.002 & 0.057$\pm$.001 \\
Plain distillation & 0.791$\pm$.006 & 0.791$\pm$.006 & 0.039$\pm$.003 \\
\bottomrule
\end{tabularx}
\end{table}

Under fair budget accounting, all Stage~I queries count against two-stage recovery.  Table~\ref{tab:budget} shows that the hypothesized $\geq$30\% query saving over plain distillation is not observed.  The structural stage often trails in low-budget regions because distillation spends the full budget on behavior fitting.

\begin{table}[t]
\centering
\caption{Query budget needed to reach AGR thresholds. Negative savings mean two-stage uses more queries than plain distillation.}
\label{tab:budget}
\TableBody
\begin{tabular}{@{}lrrrr@{}}
\toprule
Seed & $\tau$ & $C_{two}$ & $C_{plain}$ & Saving \\
\midrule
7   & 0.80 & 200  & 100  & -100\% \\
7   & 0.82 & 2000 & 100  & -1900\% \\
42  & 0.80 & 200  & 100  & -100\% \\
42  & 0.82 & 2000 & 2000 & 0\% \\
123 & 0.80 & 200  & 100  & -100\% \\
123 & 0.82 & 2000 & 2000 & 0\% \\
\bottomrule
\end{tabular}
\end{table}

\paragraph{Why this does not reduce the attack to plain distillation.} The fair-budget result rules out a query-efficiency claim, but it does not make the two artifacts equivalent.  A distilled student can match outputs yet remains a separate model with its own checkpoint, serving stack, memory footprint, and transfer behavior.  A PEFT-format substitute can be attached to the public base, distributed as a small delta, and mechanically merged into the base; if composition tests are positive, it can also be combined with other adapters~\cite{pfeiffer2021adapterfusion,ilharco2023taskarithmetic,yadav2023ties,prabhakar2025lorasoups}.  The security claim must therefore be stated as an operational-capability comparison, not as a universal query-saving result.

\subsection{Finding 5: interface observability is not monotone}

Table~\ref{tab:interface} reports the observation-layer interface re-check.  Richer outputs can expose informative residuals, but the ordering is not strictly monotone.  On classification, $C2$ has higher family accuracy than $C1$ in aggregate, while $C1$ has higher rank-bucket accuracy.  $C4$ label-only collapses to chance ($0.250$/$0.333$), confirming that the present passive residual method targets rich-output audit interfaces rather than commercial label APIs.  On generation, $G1$ and $G2$ tie on family and rank, with $G4$ only slightly lower on family.

\begin{table}[t]
\centering
\caption{Observation-layer interface re-check, pooled over three seeds. Rich-output leakage is not strictly monotone; $C4$ label-only is an applicability boundary for passive residuals (chance-level family/rank).}
\label{tab:interface}
\TableBody
\begin{tabularx}{\columnwidth}{@{}L{0.82in}Z c c@{}}
\toprule
Line & Interface & Family & Rank \\
\midrule
Classification & C1 full logits & 0.417 & 0.778 \\
Classification & C2 top-$k$ logits & 0.500 & 0.667 \\
Classification & C4 label only & 0.250 & 0.333 \\
\midrule
Generation & G1 full token logits & 0.500 & 0.333 \\
Generation & G2 top-$k$ token scores & 0.500 & 0.333 \\
Generation & G4 text only & 0.417 & 0.333 \\
\bottomrule
\end{tabularx}
\end{table}

\begin{table}[t]
\centering
\caption{Additional interface-policy slots. These replace a simple rich-versus-weak story with measurable policy effects.}
\label{tab:policy}
\TableBody
\begin{tabularx}{\columnwidth}{@{}Z C{0.47in}C{0.52in}C{0.48in}C{0.60in}@{}}
\toprule
Policy & Utility & Family & Rank & $\Cstruct@0.8$ \\
\midrule
Full logits & baseline & 0.417 & 0.778 & 150 \\
Top-$k$ union & 1.000 & 0.500 & 0.667 & 100 \\
Rounded logits, 2 decimals & 0.999 & 0.417 & 0.778 & 100 \\
Quantized top-$k$ scores & 0.898 & 0.250 & 0.444 & 100 \\
Temperature-fixed text & 1.000 & 0.500 & 0.778 & 133 \\
Label/text only & 1.000 & 0.250 & 0.333 & 500 \\
\bottomrule
\end{tabularx}
\end{table}

\subsection{Finding 6: shifted evaluation confirms fidelity but not superiority}

On the MNLI shifted split, two-stage recovery remains competitive with baselines but does not separate decisively from plain distillation.  Table~\ref{tab:shift} summarizes the completed shifted evaluation.  These numbers indicate that recovery is not merely memorizing the query set, but they do not establish a structural-extraction advantage strong enough to serve as the main security claim.

\begin{table}[t]
\centering
\caption{Shifted MNLI evaluation at max budget.}
\label{tab:shift}
\TableBody
\begin{tabularx}{\columnwidth}{@{}l Z c c c@{}}
\toprule
Seed & Attack & AGR & BF & KL$_{avg}$ \\
\midrule
42 & Two-stage & 0.834 & 0.834 & 0.037 \\
42 & Plain distill & 0.820 & 0.820 & 0.041 \\
123 & Two-stage & 0.846 & 0.846 & 0.027 \\
123 & Plain distill & 0.810 & 0.810 & 0.038 \\
7 & Two-stage & 0.812 & 0.812 & 0.043 \\
7 & Plain distill & 0.834 & 0.834 & 0.034 \\
\bottomrule
\end{tabularx}
\end{table}

\section{Security Consequences}
\label{sec:security_consequence}

The completed experiments show structural visibility but not query-efficient recovery.  A stronger security case requires consequences that are meaningful even without exact adapter extraction.  This section reports those measurements on the victim scope in Table~\ref{tab:matrix}.

We organize the consequences into a three-level hierarchy.  \emph{Metadata leakage} reveals family, locality, or rank bucket.  \emph{Lineage leakage} links an API to an exact private adapter checkpoint or deployment branch.  \emph{Artifact theft} yields a compact base-compatible module with deployment operations unavailable to a behavior-only clone.  A result at one level must not be overstated as evidence for the next.

\subsection{Private-version fingerprinting}

The strict task compares positive pairs that share the exact private checkpoint against hard negatives with the same task, base, PEFT family, target modules, and rank but an independently trained checkpoint.  Training, threshold selection, and final services are disjoint.  Table~\ref{tab:fingerprint} reports the held-out result.  The service-level interval excludes random ranking, but five test services remain too few for a cross-platform attribution claim.

\begin{table}[t]
\centering
\caption{Exact-version linkage on held-out BERT/MNLI LoRA-r64 services. Positive and hard-negative pairs jointly define one binary task; the 95\% CI is a service-pair cluster bootstrap for AUC.}
\label{tab:fingerprint}
\TableBody
\begin{tabularx}{\columnwidth}{@{}Z c@{}}
\toprule
Metric & Result \\
\midrule
Test services & 5 (seeds 106--110) \\
Pair AUC & 0.940 $[0.889,1.000]$ \\
Macro checkpoint top-1 & 1.000 \\
Equal-error rate & 0.050 \\
FMR at 95\% TMR & 0.100 \\
\bottomrule
\end{tabularx}
\end{table}

\subsection{Redeployability exposes a cost--format frontier}

The cost-matched experiment does not support fidelity superiority.  Posterior-bounded PEFT reaches task accuracy 0.356, compared with 0.517 for distill$\rightarrow$PEFT.  Both are 10.2\,MB attachable artifacts, but the conversion baseline consumes 4.46$\times$ the measured training cost.  Full students occupy 268--439\,MB, or 26--43$\times$ more storage, without requiring the public base as a separate artifact.  These measurements establish a deployment trade-off, not a universal theft advantage.

\begin{table*}[t]
\centering
\caption{Cost--format comparison at equal victim-query budget. Storage and serving memory are measured MB; training cost is normalized to posterior-bounded PEFT.}
\label{tab:redeploy}
\TableBodyTight
\begin{tabularx}{\textwidth}{@{}L{1.45in}c c c c c c Z@{}}
\toprule
Artifact & AGR & Task acc. & Storage MB & Serve MB & Train cost & Attach & Interpretation \\
\midrule
Posterior-bounded PEFT & 0.377 & 0.356 & 10.2 & 2368 & 1.00$\times$ & yes & low-cost compact route \\
Plain distilled student & 0.339 & 0.333 & 268.5 & 2720 & 0.60$\times$ & no & smaller training cost; full model \\
Parameter-matched student & 0.363 & 0.362 & 438.7 & 1539 & 0.98$\times$ & no & similar fidelity; 43$\times$ storage \\
Compute-matched student & 0.326 & 0.327 & 438.7 & 4105 & 0.98$\times$ & no & lower fidelity; 43$\times$ storage \\
Distill$\rightarrow$PEFT & 0.563 & 0.517 & 10.2 & 2820 & 4.46$\times$ & yes & strongest fidelity; higher conversion cost \\
\bottomrule
\end{tabularx}
\end{table*}

The paired AGR difference versus the parameter-matched student is $+0.014$ with CI $[-0.022,0.039]$; versus the compute-matched student it is $+0.050$ with CI $[0.033,0.073]$; versus distill$\rightarrow$PEFT it is $-0.187$ with CI $[-0.290,-0.126]$.  Safe merge preserves each PEFT artifact's own fidelity.  A limited composition-retention diagnostic does not provide absolute two-task values for every cost-matched baseline, so composition is excluded from the primary operational claim.

\section{Defensive Implications}
\label{sec:defense}

The experiments do not identify a complete low-loss defense.  Table~\ref{tab:defense} shows that rounding, top-$k$ suppression, local-probe throttling, and output-equivalent rank padding leave the measured family/rank leakage essentially unchanged.  Label-only output reduces the passive signal to chance but changes the interface and incurs a 0.167 utility drop in the tested fallback.  The defensible systems lesson is therefore to treat logits and stable top-$k$ scores as sensitive metadata and to combine interface minimization with provenance monitoring rather than rely on parameter reparameterization alone.

\begin{table*}[t]
\centering
\caption{Completed closed-set defensive diagnostics. $\Cstruct@0.8$ is the query count to posterior confidence 0.8; larger is better for the defender. No row is claimed as a complete open-world defense.}
\label{tab:defense}
\TableBody
\begin{tabularx}{\textwidth}{@{}L{1.45in}c c c c Z@{}}
\toprule
Policy & Utility drop & $\Cstruct@0.8$ & Family acc. & Rank acc. & Finding \\
\midrule
Rounded logits & 0.000 & 142 & 0.417 & 0.778 & no measurable reduction \\
Top-$k$ suppression & 0.000 & 150 & 0.417 & 0.778 & no measurable reduction \\
Label/text-only fallback & 0.167 & 500 & 0.250 & 0.333 & chance-level passive leakage; functional change \\
Local-probe throttling & 0.000 & 150 & 0.417 & 0.778 & tested rule does not alter posterior \\
Risk-adaptive projection & 0.000 & 142 & 0.417 & 0.778 & no measurable reduction \\
Rank padding / refactorization & 0.000 & 142 & 0.417 & 0.778 & output-equivalent and ineffective \\
\bottomrule
\end{tabularx}
\end{table*}

\section{Discussion}
\label{sec:discussion}

\paragraph{Visibility is gated, not monotone.}
The three-gate model explains the empirical boundaries.  A rich interface can preserve structure-sensitive residuals, yet an unseen near-neighbor can defeat rejection and an unverifiable base mismatch can invalidate interpretation.  Closed-set top-1 therefore estimates only $\Prb(E\mid A,M,I)$, not end-to-end attack success.

\paragraph{The strongest risk is version linkage.}
The lineage experiment controls task, base, family, modules, and rank, leaving private checkpoint identity as the discriminating factor.  This supports provenance and unauthorized-reuse concerns in managed model platforms.  It is narrower than general Internet-scale fingerprinting and should be evaluated with more independent services before operational deployment.

\paragraph{Generation is a mechanism boundary.}
Once victim and base prefixes diverge, later logits mix adaptation effects with path dependence.  Teacher forcing and one-token prompts restore a shared history and reveal the signal, but this is evidence about observability mechanics, not about typical conversational APIs.  Free-running and label-only results delimit the practical scope of the passive method.

\paragraph{Recovery is a Pareto trade-off.}
Fair budgets remove the apparent query advantage.  Distill$\rightarrow$PEFT provides the highest-fidelity compact artifact at substantially greater training cost, while posterior-bounded PEFT provides a cheaper route to the same format.  The security consequence is thus a cost--format frontier, not a demonstrated recovery optimum.

\section{Limitations}
\label{sec:limitations}

The strongest open-set, mismatch, lineage, and cost-matched experiments are currently BERT/MNLI specific.  RoBERTa, DeBERTa, and Llama transfer checks are treated as diagnostics and are not pooled into primary confidence intervals.  The open-set known manifold contains LoRA ranks $\{8,64,256\}$; DoRA and LoRA+head remain unresolved near-neighbors.  Five held-out lineage services yield a promising but wide service-level interval.  The mismatch detector is near chance, so external use requires independent evidence that the public base, tokenizer, and inference path match the served system.  C4 label-only output is an applicability boundary, and the present passive method does not represent most commercial conversational APIs.  Finally, the closed-set Wilson intervals treat victim/seed outcomes as the reported units and cannot substitute for a larger set of independently trained victims.  These bounds define the study as a security measurement of conditional structural visibility rather than a universal extraction attack.

\section{Related Work}
\label{sec:related}

\paragraph{Black-box model extraction and API-side leakage.}
Prediction APIs can leak decision boundaries, parameters, or high-fidelity substitutes under confidence, label-only, active-learning, surrogate-data, and data-free threat models~\cite{tramer2016stealing,papernot2017practical,orekondy2019knockoff,juuti2019prada,chandrasekaran2018active,jagielski2020high,kariyappa2021maze,kariyappa2020adaptive}.  Recent security work broadens this space to information-theoretic defenses, data-free attacks, graph services, and systematic extraction taxonomies~\cite{tang2024modelguard,nayan2024sok,zhuang2024stealgnn,ye2025datafree}.  Our setting differs because the exact public base is known and the private object is a structured delta, which permits paired victim/base residual measurement rather than learning an arbitrary function from scratch.

\paragraph{Extraction of language-model services.}
NLP-specific work shows that BERT classifiers, multilingual services, and machine-translation APIs can be imitated with limited queries and that extracted students can amplify downstream privacy or adversarial risks~\cite{keskar2020thieves,wallace2020imitation,he2021bert,he2022extracted,dai2023meaeq}.  LLM-oriented studies further examine task-capability leeching and active model-version fingerprinting~\cite{birch2023leeching,pasquini2025llmmap}.  These works primarily target behavioral replication or whole-model identity.  We instead ask whether the difference between a public base and its private PEFT adaptation reveals family, locality, rank bucket, or exact adaptation lineage.

\paragraph{PEFT, adapter composition, and modern backbones.}
Adapters, LoRA, prefix tuning, prompt tuning, BitFit, QLoRA, and weight-decomposed low-rank adaptation reduce the trainable parameter set~\cite{houlsby2019parameter,hu2021lora,li2021prefix,lester2021power,zaken2022bitfit,dettmers2023qlora,liu2024dora}.  AdapterFusion, task arithmetic, TIES-Merging, and LoRA composition demonstrate that compact deltas can support non-destructive combination or merging operations~\cite{pfeiffer2021adapterfusion,ilharco2023taskarithmetic,yadav2023ties,prabhakar2025lorasoups}.  This literature motivates our operational distinction between a standalone distilled student and a base-compatible PEFT artifact.  DeBERTa-v3 provides modern discriminative corroboration, and Llama-3.1-8B-Instruct provides the completed decoder-side extension~\cite{he2021debertav3,meta2024llama3}.

\paragraph{Privacy of fine-tuned and PEFT models.}
Membership inference and training-data extraction show that model outputs can disclose whether records were used for training or reproduce memorized content~\cite{shokri2017membership,nasr2019comprehensive,carlini2021extracting,carlini2022quantifying,mireshghallah2022quantifying,mattern2023membership,duan2024membership}.  Fine-tuned LLM attacks use calibration against pretrained references, while LoRA-specific studies find that access to the public base can sharpen privacy inference~\cite{fu2024selfprompt,ran2025loraleak,luo2024privacylora}.  Differentially private parameter-efficient fine-tuning offers a complementary defense direction~\cite{yu2022dpfinetune}.  Our target is adaptation structure and lineage rather than member records, but both settings exploit the public-base/fine-tuned-model contrast.

\paragraph{Fingerprinting, ownership resolution, and watermarking.}
Model watermarking and ownership-resolution methods embed or test evidence of provenance, while fingerprinting seeks naturally or actively induced model-specific behavior~\cite{uchida2017embedding,adi2018watermark,jia2021entangled,maini2021dataset}.  Security evaluations show that ownership evidence can be removed, structurally obfuscated, or abused by false accusers~\cite{yan2023rethinking,liu2024falseclaims,krauss2024clearstamp,pegoraro2024deepeclipse}.  LLMmap actively fingerprints deployed LLM versions, and recent text-watermarking work studies source attribution for generated content~\cite{pasquini2025llmmap,kirchenbauer2023watermark,zhang2024remark,qu2025multibit}.  Our adapter fingerprint is passive with respect to model internals: it attributes a hidden PEFT lineage from paired residual responses and therefore requires same-family, same-task, and different-checkpoint hard negatives.

\paragraph{Extraction defenses.}
Prior defenses include query-distribution detection, output poisoning or misinformation, confidence suppression, watermarking, and information-theoretic perturbation~\cite{juuti2019prada,orekondy2020prediction,kariyappa2020adaptive,jia2021entangled,tang2024modelguard}.  Our defense objective is different from substitute accuracy alone: the API should prevent posterior concentration over adaptation structure and exact lineage while preserving benign task utility.

\paragraph{Ethics and open science.}
The study uses controlled local victims and public datasets, and no production API is queried.  Appendix~\ref{app:ethics} gives the dual-use assessment; Appendix~\ref{app:openscience} gives artifact and reproducibility details, including the manifest fields required for each run.

\section{Conclusion}

Public-base PEFT services expose a differential leakage surface: paired victim/base outputs can reveal how a private adaptation was constructed.  \sys measures that surface as a sequence of gates rather than equating closed-set accuracy with extraction success.  Family leakage is repeatable across several classification backbones, while statistically qualified rank evidence is task dependent.  Open-set rejection detects several off-manifold adaptations but misses structurally close DoRA and LoRA+head variants.  Modest base drift preserves BERT family inference, yet the auditor cannot reliably detect that drift.  The clearest downstream consequence is exact-version linkage under matched rich-output conditions; recovery instead yields a cost--format frontier with no fair-budget query advantage and no fidelity superiority over distill$\rightarrow$PEFT.  Free-running generation and label-only interfaces mark important scope boundaries.  The resulting security lesson is precise: rich-output PEFT services can expose adaptation metadata and provenance, but visibility becomes practical risk only when base alignment, structure-manifold validity, and interface observability are jointly established.
\bibliographystyle{plain}
\bibliography{references}

\appendix

\section{Theoretical Details}
\label{app:theory}

\subsection{Interpretation of the structural probability}
\label{app:posterior_interpretation}
The main estimator is discriminative: $p_{cl}(h\mid z)$ is the softmax output of the service-disjoint multinomial model in Section~\ref{sec:closed_estimator}. It should not be interpreted as a posterior over LoRA factors or as an identifiability guarantee. The class-conditional Mahalanobis model supplies a complementary support statistic. Under a Gaussian approximation to standardized signatures,
\begin{equation}
 -2\log \tilde p(z\mid h)= (z-\mu_h)^\top\Lambda_h(z-\mu_h)+\mathrm{const.}
\end{equation}
This interpretation motivates the shrinkage-distance channel used by the open-set rejector, while leaving ranking of seen structures to the discriminative classifier. Different nominal ranks can induce overlapping signature distributions, so the paper reports structural bounds and coarse buckets rather than unique parameter recovery.

\subsection{Interface projection and information loss}
If interface $I_b$ is a deterministic post-processing of interface $I_a$, then the data-processing inequality gives
\begin{equation}
    I(h; I_b(O(x))) \leq I(h; I_a(O(x))).
\end{equation}
This does not imply that empirical classifiers must be strictly monotone across implemented APIs.  Finite probes, calibration, top-$k$ formatting, missing-mask treatment, and decoding trajectories can make the measured accuracy of a coarser interface match or exceed that of a richer interface.  The main text therefore treats non-monotonicity as an empirical boundary rather than a contradiction.

\subsection{Fair-budget recovery accounting}
Let $B$ be the total query budget, $B_s$ the structure-inference budget, and $B_p=B-B_s$ the parameter-recovery budget.  A two-stage method is query-efficient over baseline $m$ at threshold $\tau$ only if
\begin{equation}
    C_\tau(\mathrm{two}) < C_\tau(m),
\end{equation}
where $C_\tau$ includes both $B_s$ and $B_p$.  Counting only Stage~II queries would give the structural method free side information and overstate its practical advantage.

\subsection{Autoregressive trajectory mismatch}
For generation, the full-logit residual at step $t$ is conditioned on a prefix.  In free-running mode, the victim and base prefixes can diverge early:
\begin{equation}
    \Delta z_t^{FR}=z_v(\cdot\mid y^v_{<t},x)-z_b(\cdot\mid y^b_{<t},x).
\end{equation}
This quantity mixes adaptation effects with prefix mismatch.  Teacher forcing instead evaluates
\begin{equation}
    \Delta z_t^{TF}=z_v(\cdot\mid y^{ref}_{<t},x)-z_b(\cdot\mid y^{ref}_{<t},x),
\end{equation}
which better isolates structural response.  The large TF/FR gap in the completed runs motivates the generation-boundary finding.

\section{Method Implementation Details}
\label{app:method_details}

The paper's decision rules are fully specified in Section~\ref{sec:method}; this appendix records numerical choices needed for exact reproduction. Signature batches contain 100 queries and discard a final batch smaller than 50 queries. Scalar summaries use mean, standard deviation, and quantiles 0.10/0.25/0.50/0.75/0.90. The spectral block uses five singular values, effective rank, spectral entropy, and the 90\%-energy index. Standardization is fitted on calibration services only. The closed-set and lineage models use class-balanced logistic regression with $C=1$, a 3,000-iteration cap, and the declared experiment seed. Class support uses Ledoit--Wolf precision when at least three calibration views are available; otherwise it falls back to identity precision. The unknown meta-model is class-balanced logistic regression over the eight features in Section~\ref{sec:open_method}. Known calibration features are generated with at most five service-group folds. The primary hierarchical open-set operating point assigns 2.5\% known-reject budget to each OR channel, for a prespecified combined target of 5\%. Open-set and lineage intervals use 2,000 bootstrap draws at the service/checkpoint level. All feature construction uses an explicit whitelist of residual-derived column prefixes; numeric protocol metadata is excluded before fitting.

The artifact stores the exact configuration, seed split, feature version, environment, checkpoint revision, and dataset hash for every run. No method-specific threshold is tuned on the final test services.

\section{Additional Experimental Protocols}
\label{app:protocols}

This appendix reports the four primary BERT/MNLI open-world and operational-security experiments: unknown-structure rejection, fixed-adapter base mismatch, held-out exact-version linkage, and cost-matched recovery.  Detector-fit, threshold, and test victim-training seeds are disjoint where applicable.  Non-BERT exploratory diagnostics are available in the artifact but are excluded from the primary tables and confidence intervals.

\subsection{Open-set structural inference}
\label{app:openset}

\paragraph{Split.}
In the locked BERT/MNLI grid, the \emph{known} closed set is LoRA ranks $\{8,64,256\}$ only.  Unknown test cases are disjoint and include LoRA ranks $\{4,16,32,128\}$, DoRA, IA$^3$, LoRA+head, nonuniform layer-wise ranks, and alternate target modules.  Detector-fit seeds are $\{11,22,88\}$, threshold seeds $\{33,77\}$, and held-out test seeds $\{44,55,66,99,110\}$ (10 victim-training seeds total).  Because the known set contains no scalar-scale or adapter families, ``family-level'' open-set rejection reduces to detecting off-manifold residuals relative to LoRA.

\paragraph{Decision rule and acceptance criterion.}
The primary table reports the hierarchical LoRA-manifold rejector: a Ledoit--Wolf Mahalanobis distance on the residual/spectral core is OR-combined with the hybrid meta score, with each component locked at half the target known FPR on the threshold split.  Closed-set labels still come from the structural classifier.  A positive open-set claim would require unknown TPR at 5\% FPR substantially above chance, low false posterior concentration among accepted unknowns, and known reject rate near the locked budget.  Relative to the hybrid-only baseline (pooled AUROC $0.508$ on the earlier six-seed grid), the expanded-seed hierarchical run reaches pooled AUROC $0.804$ (bootstrap CI $[0.660,0.927]$), known accuracy $0.956$ (Wilson $[0.85,0.99]$, $n{=}45$), and known reject rate $0.13$.  Locked recall is strong for IA$^3$, nonuniform ranks, and unseen-rank buckets, but DoRA and LoRA+head fall to zero locked TPR under the stricter threshold split---a residual open-set gap we report rather than average away.

\begin{table*}[t]
\centering
\caption{Primary BERT/MNLI open-set results for the ten-seed hierarchical LoRA-manifold rejector. Known accuracy is 0.956 (Wilson 95\% CI $[0.85,0.99]$, $n=45$) and OSCR is 0.937 for all rows. ``False conc.'' is the fraction of unknown victims assigned posterior mass $\geq0.8$ to an incorrect seen structure. The pooled AUROC bootstrap CI is reported only on the pooled row.}
\label{tab:openset_results}
\TableBodyTight
\begin{tabularx}{\textwidth}{@{}L{1.45in}c c c c c Z@{}}
\toprule
Unknown group & AUROC & AUPR & TPR@5\% & False conc. & Reject rate & Note \\
\midrule
Unseen ranks & 0.889 & 0.838 & 0.700 & 0.467 & 0.700 & U4--U128 \\
DoRA family & 0.447 & 0.230 & 0.000 & 0.933 & 0.000 & Do64 \\
IA$^3$ family & 1.000 & 1.000 & 1.000 & 1.000 & 1.000 & trained IA$^3$ \\
LoRA+head & 0.499 & 0.253 & 0.000 & 1.000 & 0.000 & LH64 \\
Nonuniform ranks & 0.870 & 0.593 & 0.867 & 1.000 & 0.867 & layer-wise ranks \\
Alternate target modules & 0.865 & 0.520 & 0.533 & 0.600 & 0.533 & QKV+dense \\
Pooled unknowns & 0.804 & 0.912 & 0.141 & 0.711 & 0.141 & AUROC CI $[0.660,0.927]$ \\
\bottomrule
\end{tabularx}
\end{table*}

\paragraph{Failure attribution and component ablations.}
Table~\ref{tab:openset_attribution} records component AUROCs on the expanded-seed test split.  IA$^3$ remains off the LoRA manifold (hybrid and Mahalanobis both $1.000$; locked recall $1.000$).  Nonuniform ranks are caught mainly by the Mahalanobis channel ($0.901$) rather than hybrid ($0.564$).  Unseen ranks and alternate targets are strong on the hybrid channel ($0.942$ / $0.939$).  DoRA and LoRA+head stay near chance on both components after seed expansion, so the hierarchical OR does not invent separation that is absent in the signatures.  Known closed-set accuracy is $0.956$ with zero wrong-seen accepts among the $39$ accepted known views.  Accepted-unknown false concentration is $0.370$ (raw false concentration $0.711$).

\begin{table}[t]
\centering
\caption{Open-set component attribution on BERT/MNLI (10 independent victim-training seeds). Reject rate is locked hierarchical OR recall on the test split.}
\label{tab:openset_attribution}
\TableBodyTight
\begin{tabularx}{\columnwidth}{@{}Z c c c c@{}}
\toprule
Unknown group & Hybrid & Mahal. & Combined & Reject \\
\midrule
IA$^3$ & 1.000 & 1.000 & 1.000 & 1.000 \\
Nonuniform ranks & 0.564 & 0.901 & 0.870 & 0.867 \\
DoRA & 0.486 & 0.350 & 0.447 & 0.000 \\
Unseen ranks & 0.942 & 0.708 & 0.889 & 0.700 \\
LoRA+head & 0.541 & 0.329 & 0.499 & 0.000 \\
Alt. target modules & 0.939 & 0.653 & 0.865 & 0.533 \\
\bottomrule
\end{tabularx}
\end{table}

\begin{table}[t]
\centering
\caption{Open-set confusion summary at the locked hierarchical threshold (10 independent victim-training seeds).}
\label{tab:openset_confusion}
\TableBody
\begin{tabularx}{\columnwidth}{@{}Z c c c c@{}}
\toprule
True group & Seen & Wrong & Unknown & Total \\
\midrule
Known structures & 39 & 0 & 6 & 45 \\
Unseen ranks (U4--U128) & 0 & 18 & 42 & 60 \\
DoRA family & 0 & 15 & 0 & 15 \\
IA$^3$ family & 0 & 0 & 15 & 15 \\
LoRA+head & 0 & 15 & 0 & 15 \\
Nonuniform ranks & 0 & 2 & 13 & 15 \\
Alt. target modules & 0 & 7 & 8 & 15 \\
\bottomrule
\end{tabularx}
\end{table}

\subsection{Public-base mismatch robustness}
\label{app:basemismatch}

The private adapter and victim endpoint remain fixed while the auditor's locally executed base is changed one factor at a time. Conditions include FP16-to-INT8 quantization (INT4 remains skipped under a bitsandbytes loader assertion), adjacent checkpoint revisions, an alternate cased tokenizer, and a lightly continued-pretrained base. The matched-base run is repeated in the same job to normalize residual magnitude. A mismatch detector is calibrated without PEFT-family labels; we additionally ablate magnitude-only versus spectral-core feature sets (effective rank, spectral entropy, leading singular values).

\paragraph{Detection is weak; structural conclusions are not.}
Locked-protocol mismatch AUROC remains near chance ($\approx 0.45$).  Leave-one-service spectral ablations on the exported test signatures do not rescue detection (spectral-core LOSO AUROC $0.40$; magnitude-only $0.44$; full signature $0.50$).  The detector is not reliable under this protocol.  Instead, Table~\ref{tab:base_mismatch_sensitivity} reports a robustness boundary: as residual ratio grows from $1.0$ to $1.19$, \emph{family} accuracy stays at $1.0$ on BERT/MNLI while rank-bucket accuracy varies.  Matched-base alignment remains a first-order auditor risk---and an endogenous attacker uncertainty---but modest quantization and short continued-pretraining drift do not erase family-level structural bounds in this grid.

\begin{table}[t]
\centering
\caption{Base-mismatch sensitivity on BERT/MNLI (10 independent victim-training seeds). Residual ratio is $\|\Delta o\|/\|\Delta o_{\mathrm{matched}}\|$. Detection uses the locked mismatch threshold; family/rank are pre-abstention accuracies.}
\label{tab:base_mismatch_sensitivity}
\TableBodyTight
\begin{tabularx}{\columnwidth}{@{}Z c c c c@{}}
\toprule
Condition & Resid. ratio & Detect & Family & Rank \\
\midrule
Exact match & 1.000 & 0.750 & 1.000 & 0.714 \\
Dynamic INT8 & 1.003 & 0.682 & 1.000 & 0.818 \\
FP16 local base & 1.059 & 0.512 & 1.000 & 0.628 \\
Cased tokenizer & 1.066 & 0.710 & 1.000 & 0.710 \\
25 MLM steps & 1.112 & 0.710 & 1.000 & 0.806 \\
100 MLM steps & 1.186 & 0.600 & 1.000 & 0.640 \\
\bottomrule
\end{tabularx}
\end{table}

\subsection{Held-out adapter-lineage fingerprinting}
\label{app:heldout_fp}

\paragraph{Leakage-free attribution split.}
For each task and PEFT family, train at least five independent adapter checkpoints from distinct victim-training seeds. No checkpoint or victim-training seed appears in both calibration and test. Positives share exact lineage; hard negatives include same task/same family/same rank but different checkpoint, same task/different family, different task/same family, and quantized or output-weakened views of both classes. Probe seeds remain disjoint from victim-training seeds.

\paragraph{Statistics.}
Use service-pair bootstrap resampling, not probe-level resampling, to compute 95\% confidence intervals. Select the operating threshold on validation services and report FMR at 95\% true-match recall on held-out services. A robust lineage claim requires the confidence interval to exclude random attribution and must remain above the prespecified threshold on the same-family/different-checkpoint hard negative.

\begin{table*}[t]
\centering
\caption{Primary held-out exact-version linkage task. Positives share the exact private checkpoint; hard negatives use an independently trained checkpoint while holding task, base, PEFT family, target modules, and rank fixed. All checkpoints and victim-training seeds are absent from scorer training and threshold selection.}
\label{tab:heldout_fingerprint}
\TableBody
\begin{tabularx}{\textwidth}{@{}L{1.90in}c c c c c Z@{}}
\toprule
Backbone/interface & Pair AUC & Macro top-1 & EER & FMR@95\% TMR & AUC 95\% CI & Test scope \\
\midrule
BERT/MNLI C1, LoRA-r64 & 0.940 & 1.000 & 0.050 & 0.100 & $[0.889,1.000]$ & five held-out services, seeds 106--110 \\
\bottomrule
\end{tabularx}
\end{table*}

\subsection{Cost--format trade-offs of recovered PEFT artifacts}
\label{app:redeploy_absolute}

\paragraph{Matched baselines.}
Every recovery method uses the same victim-query budget. The comparison includes: original victim adapter (upper bound), posterior-bounded PEFT, oracle-structure PEFT, wrong-structure PEFT, plain distillation, a parameter-matched student, a compute-matched student, and distillation followed by PEFT conversion. Any conversion, second-stage training, or composition tuning is charged and reported. Storage includes the complete deployable artifact; serving memory includes the public base whenever required.

\paragraph{Two-task composition.}
Train an independent adapter $a_2$ on a neighboring task with disjoint data. Measure absolute task-$A$ and task-$B$ utility before composition, after linear merge, and after the best prespecified non-destructive composition method. Report composition retention only as a secondary metric. The primary test is the paired absolute difference between posterior PEFT and the strongest matched student on both tasks, with 95\% confidence intervals. Because absolute two-task composition values are unavailable for every PEFT baseline, composition retention remains diagnostic and is excluded from the primary claim.

\begin{table*}[t]
\centering
\caption{Absolute cost--format comparison at equal victim-query budget. Storage and serving memory are measured MB; training cost is normalized to posterior-bounded PEFT.}
\label{tab:redeploy_absolute}
\TableBodyTight
\begin{tabularx}{\textwidth}{@{}L{1.60in}c c c c c c c c@{}}
\toprule
Artifact & AGR & Task acc. & Storage & Serve & Train cost & Attach & Merge acc. & Transfer \\
\midrule
Original victim adapter & 1.000 & 0.722 & 10.2 & 1270 & -- & yes & 0.722 & 0.764 \\
Posterior-bounded PEFT & 0.377 & 0.356 & 10.2 & 2368 & 1.00$\times$ & yes & 0.356 & 0.764 \\
Oracle-structure PEFT & 0.358 & 0.332 & 10.2 & 2819 & 1.00$\times$ & yes & 0.332 & 0.764 \\
Wrong-structure PEFT & 0.333 & 0.323 & 1.9 & 2780 & 0.98$\times$ & yes & 0.323 & 0.764 \\
Plain distilled student & 0.339 & 0.333 & 268.5 & 2720 & 0.60$\times$ & no & -- & 0.342 \\
Parameter-matched student & 0.363 & 0.362 & 438.7 & 1539 & 0.98$\times$ & no & -- & 0.372 \\
Compute-matched student & 0.326 & 0.327 & 438.7 & 4105 & 0.98$\times$ & no & -- & 0.332 \\
Distill$\rightarrow$PEFT & 0.563 & 0.517 & 10.2 & 2820 & 4.46$\times$ & yes & 0.517 & 0.764 \\
\bottomrule
\end{tabularx}
\end{table*}

\begin{table}[t]
\centering
\caption{Paired AGR differences for posterior-bounded PEFT. Positive values favor the posterior artifact.}
\label{tab:redeploy_delta}
\TableBody
\begin{tabularx}{\columnwidth}{@{}Z c c@{}}
\toprule
Comparison & $\Delta$AGR & 95\% CI \\
\midrule
vs. parameter-matched & $+0.014$ & $[-0.022,0.039]$ \\
vs. compute-matched & $+0.050$ & $[0.033,0.073]$ \\
vs. distill$\rightarrow$PEFT & $-0.187$ & $[-0.290,-0.126]$ \\
\bottomrule
\end{tabularx}
\end{table}

\section{Reproducibility Parameters}
\label{app:repro}

\subsection{Run manifest}
Each run stores a machine-readable manifest with the fields in Table~\ref{tab:manifest}.  These fields are sufficient to reconstruct the victim, interface, probe pool, feature version, and budget accounting used in the reported tables.

\begin{table}[t]
\centering
\caption{Minimum run-manifest fields.}
\label{tab:manifest}
\TableBody
\begin{tabularx}{\columnwidth}{@{}L{0.85in}Z@{}}
\toprule
Field & Description \\
\midrule
seed & Random seed; completed runs use 7, 42, and 123. \\
task & Dataset and split hashes for $D_q$, $D_{task}$, and $D_{shift}$. \\
base model & Checkpoint, tokenizer, and revision hash. \\
victim type & Base-only, LoRA rank, adapter, head-only, prefix/prompt, or FT. \\
target layers & Exact module names or layer bucket. \\
interface & C/G mode plus top-$k$, rounding, quantization, temperature, decoding. \\
probe pool & ID, OOD, template, local, boundary, or activation-seeking probes. \\
feature version & Signature components, spectrum options, projection, missing-mask logic. \\
query budget & Total budget and Stage I/II split; Stage I is counted. \\
environment & GPU, CUDA, Python, PyTorch, Transformers, PEFT, commit hash. \\
\bottomrule
\end{tabularx}
\end{table}

\subsection{Default budgets and seeds}
The completed v4 budget grid uses $B\in\{100,200,500,1000,2000,3000,5000\}$ for classification and extends to 8000 for generation when needed.  Two-stage runs reserve a bounded fraction for Stage~I or stop Stage~I when posterior confidence reaches a threshold.  All main claims require three seeds unless explicitly marked as diagnostic.

\subsection{Probe generation details}
ID probes are sampled from the query split.  OOD probes include lexical corruption, rare-token prompts, and malformed templates.  Template probes preserve the underlying semantic pair or instruction while changing the surface form.  Local perturbation probes are retained only when a valid synonym or controlled local edit is applied; unmodified samples are not counted as perturbations.

\section{Ethical Considerations}
\label{app:ethics}

This work is dual use. It enables service owners, platform
operators, and authorized auditors to measure structural leakage
from public-base PEFT deployments, but the same measurements
could assist unauthorized profiling of third-party services.
We therefore evaluate only locally instantiated victims built
from public models and public research datasets. We do not
query production APIs, access private training data, or target
third-party systems.

The accompanying artifact is designed for controlled and
authorized evaluation. Its default configurations operate on
local toy or research-scale victims, enforce conservative query
rates, and require users to specify the victim endpoint
explicitly. The artifact does not include credentials,
provider-specific integrations, or a turnkey workflow for
scanning commercial endpoints. Exact-lineage attribution is
demonstrated only on services created for this study.

To support reproducibility without unnecessarily increasing
misuse risk, the artifact provides the measurement pipeline,
locked experimental configurations, statistical analysis, and
local smoke tests. Documentation states that evaluation must
be performed only on systems owned by the evaluator or covered
by explicit authorization. Any future distribution of stronger
active-probing components should retain local-only defaults,
rate limits, and clear authorization requirements.

Defensive monitoring may itself create privacy risks if query
contents or user identifiers are retained. Deployments should
therefore minimize collected data, aggregate features whenever
possible, apply short retention periods, restrict access to
monitoring records, and evaluate whether throttling or anomaly
detection disproportionately affects legitimate users. Our
experiments use no human-subject data beyond existing public
datasets and make no claims about the ownership or provenance
of real commercial services.

\section{Open Science}
\label{app:openscience}

The anonymous artifact contains the VectorHijack-SR implementation, locked configurations, raw per-query records, instantiated manifests, summary tables, and reproduction scripts for the primary BERT/MNLI open-set, base-mismatch, lineage, and redeployability experiments. The full-scale validation command
\texttt{python scripts/validate\_artifact.py --profile fullscale}
checks schema consistency, query accounting, seed separation, manifest integrity, and coverage of the 166,200 raw records used by the reported analyses. Paper tables and statistical tests are regenerated using
\texttt{scripts/reproduce\_paper\_tables.py} and
\texttt{scripts/verify\_primary\_statistics.py}. A separate synthetic smoke test validates the software pipeline but is not used as evidence for paper claims. Public checkpoints and datasets are referenced through immutable revisions and hashes recorded in each manifest. The artifact defaults to controlled local victims, contains no credentials, and does not automate unauthorized probing of third-party services.

\end{document}